\documentclass[11pt]{article}
\usepackage{amsfonts}
\usepackage{amssymb,amsmath,mathrsfs,mathdots}
\usepackage{graphicx,subfigure}
\usepackage{indentfirst}
\usepackage{cite}

\begin{document}
\title{Darboux transformation and soliton solutions of the generalized Sasa-Satsuma equation}
\author{Hong-Qian Sun and Zuo-Nong Zhu
\footnote{Author to whom correspondence should be addressed: znzhu@sjtu.edu.cn}\\
School of Mathematical Sciences, Shanghai Jiao Tong University,\\ 800 Dongchuan Road, Shanghai, 200240, P. R. China}
\date{}
\maketitle
\begin{abstract}
The Sasa-Satsuma equation, a higher-order nonlinear Schr\"{o}dinger equation, is an important integrable equation, which displays the propagation of femtosecond pulses in optical fibers. In this paper, we investigate a generalized Sasa-Satsuma(gSS) equation. The Darboux transformation(DT) for the focusing and defocusing gSS equation is constructed. By using the DT, various of soliton solutions for the generalized Sasa-Satsuma equation are derived, including hump-type, breather-type and periodic soliton. Dynamics properties and asymptotic behavior of these soliton solutions are analyzed. Infinite number conservation laws and conserved quantities for the gSS equation are obtained.\\
{\bf keyword:} The generalized Sasa-Satsuma equation, ~~Darboux transformation,~~Breather-type soliton solutions,~~ Asymptotic behavior of soliton solutions,~~ Infinite number conservation laws
\end{abstract}
\section{Introduction}

The nonlinear Schr\"{o}dinger(NLS) equation is an important integrable equation, which is also a fundamental equation in nonlinear physics, which describes soliton propagation in nonlinear fiber optics, water waves, plasma physics, etc. The Sasa-Satsuma equation
\begin{equation}\label{SaSa}
\text{i}q_{_T}+\frac{\epsilon}{2}q_{_{XX}}+q|q|^2+\text{i}\left(q_{_{XXX}}+3\epsilon(2q_{_X}|q|^2+q(|q|^2)_{_X})\right)=0,
\end{equation}
a higher-order NLS equation, displays the propagation of femtosecond pulses in optical fibers. Here $\epsilon=1$ and $\epsilon=-1$ display the focusing case and defocusing case, respectively.
Under the variable transformations
\begin{equation}\label{VT1}
u(x,t)=q(X,T)\text{exp}\left\{-\frac{\text{i}\epsilon}{6}(X-\frac{T}{18})\right\},~t=T,~x=X-\frac{T}{12},
\end{equation}
Eq.(1) changes into a complex modified KdV-type equation
\begin{equation}\label{SS}
u_t+u_{xxx}+3\epsilon(2|u|^2u_x+u(|u|^2)_x)=0.
\end{equation}
The Sasa-Satsuma equation has been extensively studied in the several topic, e.g. Cauchy problem by the inverse scattering transform(IST)[1-3], soliton solutions, including bright-soliton, dark-soliton, double-hump, breather soliton, resonant $2$-solitons solution, rogue wave and $W$-shape soliton, by the DT method and Hirota's bilinear method[4-11], the initial-boundary value problem by the Fokas method[12,13], long-time asymptotic by the nonlinear steepest descent method[14,15].

In this paper, we investigate a generalized Sasa-Satsuma(gSS) equation, introduced in [16]
\begin{equation}\label{gSS}
u_t+u_{xxx}-3\epsilon(2a|u|^2u_x+2bu^2u_x+au(|u|^2)_x+bu^{*}(|u|^2)_x)=0,
\end{equation}
where $a,b$ are real constants satisfying $|a|\neq|b|$ and $^*$ represents the complex conjugate.
We remark here that under the variable transformation (2), the gSS equation (4) becomes into
\begin{equation}\label{gSaSa}\begin{gathered}
\text{i}q_{_T}+\frac{\epsilon}{2}q_{_{XX}}-aq|q|^2+\text{i}q_{_{XXX}}-3\text{i}\epsilon a(2q_{_X}|q|^2+q(|q|^2)_{_X})\\
-3\text{i}\epsilon b\text{e}^{\frac{\text{i}\epsilon}{3}(X-\frac{T}{18})}\left(\text{e}^{-\frac{2\text{i}\epsilon}{3}(X-\frac{T}{18})}(2q^2q_X-\frac{\text{i}\epsilon}{3}q^3)+|q|^2q_x^*+q^{*2}q_x\right)=0.
\end{gathered}\end{equation}
Clearly, equation (5) with $a=-1,b=0$ is just Sasa-Satsuma equation (3). In this sense, equation (4) is really a generalized Sasa-Satsuma equation, and then the research on equation (4) is important for nonlinear optics.
For the focusing gSS equation (4), its soliton solutions were obtained by using the Riemann-Hilbert approach[16]; the long-time asymptotic behavior of Eq. (4) was discussed [17]. To the best of our knowledge, the soliton solutions with non-zero seed solution for the gSS equation (4) have not been studied.

As we known, the Darboux transformation is an very important method for solving an integrable equation. But, the construction of DT is difficult. In this paper, our main purpose is to construct DT for the gSS equation (4). And then, by using our DT, various of soliton solutions for the gSS equation are derived, including hump-type, breather-type and periodic soliton. With the zero seed solution, we obtain single-, double-hump soliton, single-, double-peak breather solution to the focusing gSS equation. Based on the nonzero seed solution, we get periodic soliton, bright-dark breather, bright-bright breather, resonant $2$-breather solution. Furthermore, dynamics properties and asymptotic behavior of these solutions are analyzed. The infinite number conservation laws for the gSS equation are obtained.


\section{ \bf The construction of DT for the gSS equation}

In this section, $N$-fold DT of the gSS equation is constructed.

The Lax pair of the gSS equation (4) is given(see [16]) by
\begin{equation}\label{gSSLP}\begin{gathered}
\Psi_x=U(\lambda,Q)\Psi,\quad\Psi_t=V(\lambda,Q)\Psi,\\
U(\lambda,Q)=\text{i}\lambda\Lambda+Q,V(\lambda,Q)=4\text{i}\lambda^3\Lambda+4\lambda^2Q+2\text{i}\lambda(Q^2+Q_x)\Lambda+Q_xQ-QQ_x-Q_{xx}+2Q^3,
\end{gathered}\end{equation}
where $\Psi$ is a matrix function, $\lambda$ is the spectral parameter, and
\begin{equation}\begin{gathered}
Q=\left(
\begin{array}{ccc}
0&0&u\\
0&0&\epsilon u^{*}\\
\epsilon(au^{*}+bu)&au+bu^{*}&0
\end{array}
\right), \Lambda=\text{diag}(1,1,-1).
\end{gathered}\end{equation}

Suppose that $|y_{j}\rangle=(\psi_{1}^{(j)},\psi_{2}^{(j)},\psi_{3}^{(j)})^T$ is an eigenfunction for the eigenvalue problem (6) at $\lambda=\lambda_j$, then $|\eta_{j}\rangle=(\psi_{2}^{(j)^{*}},\psi_{1}^{(j)^{*}},\epsilon\psi_{3}^{(j)^{*}})^{T}$ is also an eigenfunction of the eigenvalue problem (6) at $\lambda=-\lambda_j^{*}$, and $|\theta_{j}\rangle=\langle y_{j}|J$ is a solution to the adjoint problem of Eq.(4)
\begin{equation}\label{adjoint}
|\theta\rangle_x=-\theta U,\quad |\theta\rangle_t=-\theta V,
\end{equation}
at $\lambda=\lambda_j^*$, where
\begin{equation*}
\langle y_{j}|=|y_{j}\rangle^{\dag},~~~J=\left(
\begin{array}{ccc}
-a\epsilon&-b&0\\
-b&-a\epsilon&0\\
0&0&1
\end{array}
\right),
\end{equation*}
and $\dag$ represents the complex conjugate transpose.
By constructing the DT for the gSS equation (4), we obtain our main result.

{\bf Theorem 1.} Under a gauge transform $\Psi^{(1)}=T^{(1)}\Psi$, where
\begin{equation}\label{DT0}
T^{(1)}=I-(|y_{1}\rangle,|\eta_{1}\rangle)\left(
\begin{array}{cc}
\frac{\langle y_{1}|J|y_{1}\rangle}{\lambda_1^*-\lambda_1}& \frac{\langle y_{1}|J|\eta_{1}\rangle}{2\lambda_1^*}\\
\frac{\langle \eta_{1}|J|y_{1}\rangle}{-2\lambda_1}& \frac{\langle \eta_{1}|J|\eta_{1}\rangle}{-\lambda_1+\lambda_1^*}
\end{array}
\right)^{-1}\left(
\begin{array}{cc}
\frac{\langle y_{1}|J}{\lambda_1^*-\lambda}\\ \frac{\langle \eta_{1}|J}{-\lambda_1-\lambda}
\end{array}
\right)\triangleq I-\mathbb{K}_1W_1^{-1}\Gamma(\mathbb{K}_1).
\end{equation}
one can find that the eigenvalue problem (6) changes into
\begin{equation}\label{problem0}
\Psi_x^{(1)}=U^{(1)}(\lambda,Q^{(1)})\Psi^{(1)},~~~~\Psi_t^{(1)}=V^{(1)}(\lambda,Q^{(1)})\Psi^{(1)},
\end{equation}
where
\begin{equation}\label{relation1}
Q^{(1)}=Q+\text{i}\left[\mathbb{K}_1W_1^{-1}\mathbb{K}_1^{\dag}J,\Lambda\right].
\end{equation}
We could conclude that $T^{(1)}$ is DT of the spectral problem (6), and the relation between the old and new solution of the gSS equation (4) can be written as
\begin{equation}\label{relation2}
u^{(1)}=u-\frac{2\text{i}}{A_1^2+|B_1|^2}\left(A_1(\psi_{1}^{(1)}\psi_{3}^{(1)^*}+\epsilon\psi_{2}^{(1)^*}\psi_{3}^{(1)})-\epsilon B_1\psi_{1}^{(1)}\psi_{3}^{(1)}+B_1^*\psi_{2}^{(1)^*}\psi_{3}^{(1)^*}\right),
\end{equation}
where $A_1=\frac{\langle y_{1}|J|y_{1}\rangle}{\lambda_1^*-\lambda_1},B_1=\frac{\langle y_{1}|J|\eta_{1}\rangle}{2\lambda_1^*}$.\\
{\bf Proof.} It is obvious that we have
\begin{equation*}
\Lambda J=J\Lambda,~~Q^{\dag}J=-JQ,~~\langle \eta_{1}|J|\eta_{1}\rangle=\langle y_{1}|J|y_{1}\rangle,~~\langle \eta_{1}|J|y_{1}\rangle=(\langle y_{1}|J|\eta_{1}\rangle)^*.
\end{equation*}
Then we can verify the following equations
\begin{equation*}
(\langle y_{1}|J|y_{1}\rangle)_x=-\text{i}(\lambda_1^*-\lambda_1)\langle y_{1}|J\Lambda|y_{1}\rangle,~~
(\langle y_{1}|J|\eta_{1}\rangle)_x=-2\text{i}\lambda_1^*\langle y_{1}|J\Lambda|\eta_{1}\rangle.
\end{equation*}

With a direct calculation, we have
\begin{equation}\label{KW}\begin{aligned}
&\mathbb{K}_1^{\dag}J=(D_1^{\dag}-\lambda I)\Gamma(\mathbb{K}_1),~~ D_1^{\dag}W_1-W_1D_1=\mathbb{K}_1^{\dag}J\mathbb{K}_1,~~\mathbb{K}_{1,x}=\text{i}\Lambda \mathbb{K}_1D_1+Q\mathbb{K}_1,\\ \nonumber
&W_{1,x}=-\text{i}\mathbb{K}_1^{\dag}J\Lambda \mathbb{K}_1,~~(\Gamma(\mathbb{K}_1)\Psi)_x=-\text{i}\mathbb{K}_1^{\dag}J\Lambda\Psi,~~D_1=\text{diag}(\lambda_1,-\lambda_1^*),\nonumber
\end{aligned}\end{equation}
and then we obtain
\begin{equation}\begin{aligned}
&\Psi_x^{(1)}=(\text{i}\lambda\Lambda+Q)\Psi-\text{i}\Lambda \mathbb{K}_1D_1W_1^{-1}\Gamma(\mathbb{K}_1)\Psi-Q\mathbb{K}_1W_1^{-1}\Gamma(\mathbb{K}_1)\Psi+\text{i}\mathbb{K}_{1}W_1^{-1}\mathbb{K}_1^{\dag}J\Lambda\Psi\\
&\quad\quad~~~-\text{i}\mathbb{K}_{1}W_1^{-1}\mathbb{K}_1^{\dag}J\Lambda\mathbb{K}_1W_1^{-1}\Gamma(\mathbb{K}_1)\Psi\\
&\quad\quad=(\text{i}\lambda\Lambda+Q)\Psi-\text{i}\lambda\Lambda \mathbb{K}_1W_1^{-1}\Gamma(\mathbb{K}_1)\Psi+\text{i}\Lambda \mathbb{K}_1W_1^{-1}\mathbb{K}_1^{\dag}J\mathbb{K}_1W_1^{-1}\Gamma(\mathbb{K}_1)\Psi\\
&\quad\quad~~~-\text{i}\Lambda \mathbb{K}_1W_1^{-1}\mathbb{K}_1^{\dag}J\Psi-Q\mathbb{K}_1W_1^{-1}\Gamma(\mathbb{K}_1)\Psi+\text{i}\mathbb{K}_{1}W_1^{-1}\mathbb{K}_1^{\dag}J\Lambda\Psi\\
&\quad\quad~~~-\text{i}\mathbb{K}_{1}W_1^{-1}\mathbb{K}_1^{\dag}J\Lambda\mathbb{K}_1W_1^{-1}\Gamma(\mathbb{K}_1)\Psi\\
&\quad\quad=(\text{i}\lambda\Lambda+Q+\text{i}\mathbb{K}_{1}W_1^{-1}\mathbb{K}_1^{\dag}J\Lambda-\text{i}\Lambda \mathbb{K}_{1}W_1^{-1}\mathbb{K}_1^{\dag}J)(\Psi-\mathbb{K}_{1}W_1^{-1}\Gamma(\mathbb{K}_1)\Psi).
\end{aligned}\end{equation}

Next, let us prove that $Q^{(1)}$ has the same structure with $Q$. Denote $\Theta=\mathbb{K}_1W_1^{-1}\mathbb{K}_1^{\dag}J$,
and then Eq. (11) can be rewritten as
\begin{equation*}
Q^{(1)}=\left(
\begin{array}{ccc}
0&0&u-2\text{i}\Theta_{13}\\
0&0&\epsilon u^{*}-2\text{i}\Theta_{23}\\
\epsilon(au^{*}+bu)+2\text{i}\Theta_{31}&au+bu^{*}+2\text{i}\Theta_{32}&0
\end{array}
\right),
\end{equation*}
where
\begin{equation}\begin{aligned}
&\Theta_{13}=\frac{1}{A_1^2+|B_1|^2}(\psi_1,\psi_2^*)
\left(
\begin{array}{cc}
A_1&-B_1\\ B_1^*&A_1
\end{array}
\right)\left(
\begin{array}{cc}
\psi_3^*\\ \epsilon \psi_3
\end{array}
\right),\\
&\Theta_{23}=\frac{1}{A_1^2+|B_1|^2}(\psi_2,\psi_1^*)
\left(
\begin{array}{cc}
A_1&-B_1\\ B_1^*&A_1
\end{array}
\right)\left(
\begin{array}{cc}
\psi_3^*\\ \epsilon \psi_3
\end{array}
\right),\\
&\Theta_{31}=\frac{1}{A_1^2+|B_1|^2}(\psi_3,\epsilon\psi_3^*)
\left(
\begin{array}{cc}
A_1&-B_1\\ B_1^*&A_1
\end{array}
\right)\left(
\begin{array}{cc}
-a\epsilon\psi_1^*-b\psi_2^*\\ -a\epsilon\psi_2-b\psi_1
\end{array}
\right),\\
&\Theta_{31}=\frac{1}{A_1^2+|B_1|^2}(\psi_3,\epsilon\psi_3^*)
\left(
\begin{array}{cc}
A_1&-B_1\\ B_1^*&A_1
\end{array}
\right)\left(
\begin{array}{cc}
-b\psi_1^*-a\epsilon\psi_2^*\\ -b\psi_2-a\epsilon\psi_1
\end{array}
\right).
\end{aligned}\end{equation}
It is easy to prove that
\begin{equation}
\Theta_{23}=-\epsilon\Theta_{13}^*,~~\Theta_{31}=\epsilon(a\Theta_{13}^*-b\Theta_{13}),
~~\Theta_{32}=-a\Theta_{13}+b\Theta_{13}^*.
\end{equation}
This means that $Q^{(1)}$ has the same structure with $Q$. So, we have shown that the matrix $U^{(1)}(\lambda,Q^{(1)})$ has the same structure with $U(\lambda,Q)$.

Next, we hope to prove that the matrix $V^{(1)}(\lambda,Q^{(1)})$ has the same structure with $V(\lambda,Q)$. For the time development part, we can obtain the following equations via a long but direct calculation
\begin{equation}\begin{aligned}
&(\langle y_{1}|J|y_{1}\rangle)_t=-4\text{i}(\lambda_1^{*3}-\lambda_1^3)\langle y_{1}|J\Lambda|y_{1}\rangle-4(\lambda_1^{*2}-\lambda_1^2)\langle y_{1}|JQ|y_{1}\rangle\\
&\qquad\qquad\qquad\qquad-2\text{i}(\lambda_1^{*}-\lambda_1)\langle y_{1}|J(Q^2+Q_x)\Lambda|y_{1}\rangle,\\
&(\langle y_{1}|J|\eta_{1}\rangle)_t=-8\text{i}\lambda_1^{*3}\langle y_{1}|J\Lambda|\eta_{1}\rangle-4\text{i}\lambda_1^{*}\langle y_{1}|J(Q^2+Q_x)\Lambda|\eta_{1}\rangle,
\end{aligned}\end{equation}
and
\begin{equation}\label{KWt}\begin{aligned}
&\mathbb{K}_{1,t}=4\text{i}\Lambda \mathbb{K}_1D_1^3+4Q\mathbb{K}_1D_1^2+2\text{i}(Q^2+Q_x)\Lambda\mathbb{K}_1D_1+(Q_xQ-QQ_x-Q_{xx}+2Q^3)\mathbb{K}_1,\\ \nonumber
&W_{1,t}=-4\text{i}D_1^{_{\dag}2}\mathbb{K}_1^{\dag}J\Lambda \mathbb{K}_1-4\text{i}\mathbb{K}_1^{\dag}J\Lambda \mathbb{K}_1D_1^2-4\text{i}D_1^{\dag}\mathbb{K}_1^{\dag}J\Lambda \mathbb{K}_1D_1-4D_1^{\dag}\mathbb{K}_1^{\dag}JQ\mathbb{K}_1\\ \nonumber
&~~~~~~~~-4\mathbb{K}_1^{\dag}JQ\mathbb{K}_1D_1-2\text{i}\mathbb{K}_1^{\dag}J(Q^2+Q_x)\Lambda \mathbb{K}_1,\\ \nonumber
&(\Gamma(\mathbb{K}_1)\Psi)_t=-4\text{i}\lambda^2\mathbb{K}_1^{\dag}J\Lambda\Psi-4\text{i}\lambda D_1^{\dag}\mathbb{K}_1^{\dag}J\Lambda\Psi-4\text{i}D_1^{_{\dag}2}\mathbb{K}_1^{\dag}J\Lambda\Psi-4\lambda\mathbb{K}_1^{\dag}JQ\Psi\\ \nonumber
&~~~~~~~~-4D_1^{\dag}\mathbb{K}_1^{\dag}JQ\Psi-2\text{i}\mathbb{K}_1^{\dag}J(Q^2+Q_x)\Lambda\Psi,
\end{aligned}\end{equation}
By using equations (11), (13) and (17), we have
\begin{equation}\begin{aligned}
&\Psi_t^{(1)}=(4\text{i}\lambda^3\Lambda+4\lambda^2Q+2\text{i}\lambda(Q^2+Q_x)\Lambda+Q_xQ-QQ_x-Q_{xx}+2Q^3)\Psi\\
&~~~-4\text{i}\Lambda \mathbb{K}_1D_1^3W_1^{-1}\Gamma(\mathbb{K}_1)\Psi-(4Q\mathbb{K}_1D_1^2+2\text{i}(Q^2+Q_x)\Lambda\mathbb{K}_1D_N\\
&~~~+(Q_xQ-QQ_x-Q_{xx}+2Q^3)\mathbb{K}_1)W_1^{-1}\Gamma(\mathbb{K}_1)\Psi-\mathbb{K}_1W_1^{-1}(4\text{i}D_N^{_{\dag}2}\mathbb{K}_1^{\dag}J\Lambda \mathbb{K}_1\\
&~~~+4\text{i}\mathbb{K}_1^{\dag}J\Lambda \mathbb{K}_1D_N^2+4\text{i}D_N^{\dag}\mathbb{K}_1^{\dag}J\Lambda \mathbb{K}_1D_N+4D_N^{\dag}\mathbb{K}_1^{\dag}JQ\mathbb{K}_1+4\mathbb{K}_1^{\dag}JQ\mathbb{K}_1D_N\\
&~~~+2\text{i}\mathbb{K}_1^{\dag}J(Q^2+Q_x)\Lambda \mathbb{K}_1)W_1^{-1}\Gamma(\mathbb{K}_1)\Psi+\mathbb{K}_1W_1^{-1}(4\text{i}\lambda^2\mathbb{K}_1^{\dag}J\Lambda\Psi\\
&~~~+4\text{i}\lambda D_N^{\dag}\mathbb{K}_1^{\dag}J\Lambda\Psi+4\text{i}D_N^{_{\dag}2}\mathbb{K}_1^{\dag}J\Lambda\Psi+4\lambda\mathbb{K}_1^{\dag}JQ\Psi+4D_N^{\dag}\mathbb{K}_1^{\dag}JQ\Psi+2\text{i}\mathbb{K}_1^{\dag}J(Q^2+Q_x)\Lambda\Psi)\\
&~~~=(4\text{i}\lambda^3\Lambda+4\lambda^2(Q+\text{i}\mathbb{K}_1
W_1^{-1}\mathbb{K}_1^{\dag}J\Lambda-\text{i}\Lambda\mathbb{K}_1
W_1^{-1}\mathbb{K}_1^{\dag}J)+2\text{i}\lambda((Q^{^2}+Q_x)\Lambda\\
&~~~+2\text{i}Q\mathbb{K}_1W_1^{-1}\mathbb{K}_1^{\dag}J-2\text{i}\mathbb{K}_1W_1^{-1}\mathbb{K}_1^{\dag}JQ-2\Lambda\mathbb{K}_1D_NW_1^{-1}\mathbb{K}_1^{\dag}J+2\mathbb{K}_1W_1^{-1}D_N^{\dag}\mathbb{K}_1^{\dag}J\Lambda\\
&~~~-2\mathbb{K}_1W_1^{-1}\mathbb{K}_1^{\dag}J\Lambda\mathbb{K}_1W_1^{-1}\mathbb{K}_1^{\dag}J)+Q_xQ-QQ_x-Q_{xx}+2Q^{^3}-2\text{i}(Q^2+Q_x)\Lambda\mathbb{K}_1W_1^{-1}\mathbb{K}_1^{\dag}J\\
&~~~+2\text{i}\mathbb{K}_1W_1^{-1}\mathbb{K}_1^{\dag}J(Q^2+Q_x)\Lambda-4Q\mathbb{K}_1D_NW_1^{-1}\mathbb{K}_1^{\dag}J+4\mathbb{K}_1W_1^{-1}D_N^{\dag}\mathbb{K}_1^{\dag}JQ\\
&~~~-4\mathbb{K}_1W_1^{-1}\mathbb{K}_1^{\dag}JQ\mathbb{K}_1W_1^{-1}\mathbb{K}_1^{\dag}J-4\text{i}\Lambda\mathbb{K}_1
D_N^2W_1^{-1}\mathbb{K}_1^{\dag}J+4\text{i}\mathbb{K}_1W_1^{-1}D_N^{_{\dag}2}\mathbb{K}_1^{\dag}J\Lambda\\
&~~~-4\text{i}\mathbb{K}_1W_1^{-1}\mathbb{K}_1^{\dag}J\Lambda K_ND_NW_1^{-1}\mathbb{K}_1^{\dag}J-4\text{i}\mathbb{K}_1W_1^{-1}D_N^{\dag}\mathbb{K}_1^{\dag}J\Lambda K_NW_1^{-1}\mathbb{K}_1^{\dag}J)\\
&~~~\cdot(\Psi-\mathbb{K}_1W_1^{-1}\Gamma(\mathbb{K}_1)\Psi)\\
&~~~=(4\text{i}\lambda^3\Lambda+4\lambda^2Q^{(1)}+2\text{i}\lambda(Q^{(1)^2}+Q_x^{(1)})\Lambda+Q_x^{(1)}Q^{(1)}-Q^{(1)}Q_x^{(1)}-Q^{(1)}_{xx}+2Q^{(1)^3})\Psi^{(1)}.
\end{aligned}\end{equation}
This completes the proof of Theorem 1.

Assume that $|y_{j}\rangle=(\psi_{1}^{(j)},\psi_{2}^{(j)},\psi_{3}^{(j)})^T$($j=1,2,...N,N\geq 2$) are eigenfunctions for the eigenvalue problem (6) at $\lambda=\lambda_j$, respectively, we can construct the $N$-fold DT as the following Theorem.
\\
{\bf Theorem 2.} Take gauge transform $\Psi^{(N)}=T^{(N)}\Psi$, where $T^{N}=I-\mathbb{K}_NW_N^{-1}\Gamma(\mathbb{K}_N)$,
\begin{equation}\label{DT}\begin{aligned}
&\mathbb{K}_N=(|y_{1}\rangle,|\eta_{1}\rangle,|y_{2}\rangle,|\eta_{2}\rangle,\cdots|y_{N}\rangle,|\eta_{N}\rangle)\triangleq(K_{1},K_{2},\cdots,K_{N}),\\ \nonumber
&W_N=\left(
\begin{array}{cc}
\Omega(K_{1},K_{1})& \Omega(K_{1},K_{2}) \quad\cdots \quad\Omega(K_{1},K_{N})\\
\Omega(K_{2},K_{2})& \Omega(K_{2},K_{2}) \quad\cdots\quad \Omega(K_{2},K_{N})\\
\vdots &\vdots\qquad \ddots\qquad \vdots \\
\Omega(K_{N},K_{1})& \Omega(K_{N},K_{2}) \quad\cdots \quad\Omega(K_{N},K_{N})
\end{array}
\right),\Gamma(\mathbb{K}_N)=\left(
\begin{array}{cc}
\Gamma(K_{1})\\ \Gamma(K_{2})\\ \vdots\\ \Gamma(K_{N})
\end{array}
\right),\\ \nonumber
&\Omega(K_{i},K_{j})=\left(
\begin{array}{cc}
\frac{\langle y_{i}|J|y_{j}\rangle}{\lambda_i^*-\lambda_j}& \frac{\langle y_{i}|J|\eta_{j}\rangle}{\lambda_i^*+\lambda_j^*}\\
\frac{\langle \eta_{i}|J|y_{j}\rangle}{-\lambda_i-\lambda_j}& \frac{\langle \eta_{i}|J|\eta_{j}\rangle}{-\lambda_i+\lambda_j^*}
\end{array}
\right),\Gamma(K_{i})=\left(
\begin{array}{cc}
\frac{\langle y_{i}|J}{\lambda_i^*-\lambda}\\ \frac{\langle \eta_{i}|J}{-\lambda_i-\lambda}
\end{array}
\right), 1\leq i,j\leq N.
\end{aligned}\end{equation}
Then the eigenvalue problem (6) changes into
\begin{equation}\label{problem1}
\Psi_x^{(N)}=U^{(N)}(\lambda,Q^{(N)})\Psi^{(N)},~~~~\Psi_t^{(N)}=V^{(N)}(\lambda,Q^{(N)})\Psi^{(N)},
\end{equation}
where
\begin{equation}\label{relation11}
Q^{(N)}=Q+\text{i}\left[\mathbb{K}_NW_N^{-1}\mathbb{K}_N^{\dag}J,\Lambda\right].
\end{equation}
We have the conclusion that matrix $U^{(N)}(\lambda,Q^{(N)})$ and $V^{(N)}(\lambda,Q^{(N)})$ have the same structures with matrix $U(\lambda,Q)$ and $V(\lambda,Q)$. This means that $u(x,t)$ is a solution of the gSS equation (4) (corresponding to eigenfunction $\psi$), then $u^{(N)}(x,t)$ is also a solution of the gSS equation (4) (corresponding to eigenfunction $\psi^{(N)}$), where
\begin{equation}\label{relation12}
u^{(N)}=u-2\text{i}\mathbf{h}_1W_N^{-1}\mathbf{h}_3^{\dag},
\end{equation}
with
\begin{equation*}
\mathbf{h}_1=(\psi_{1}^{(1)},\psi_{2}^{(1)^*},\psi_{1}^{(2)},\psi_{2}^{(2)^*},\cdots,\psi_{1}^{(N)},\psi_{2}^{(N)^*}),\mathbf{h}_3=(\psi_{3}^{(1)},\epsilon\psi_{3}^{(1)^*},\psi_{3}^{(2)},\epsilon\psi_{3}^{(2)^*},\cdots,\psi_{3}^{(N)},\epsilon\psi_{3}^{(N)^*}).
\end{equation*}
{\bf Proof.} Similar to the proof of Theorem 1, we can prove that matrix $U^{(N)}(\lambda,Q^{(N)})$ and $V^{(N)}(\lambda,Q^{(N)})$ have the same structures with matrix $U(\lambda,Q)$ and $V(\lambda,Q)$. Here we only proof that $Q^{(N)}$ has the same structure with $Q$. Setting $\Theta=\mathbb{K}_NW_N^{-1}\mathbb{K}_N^{\dag}J$, Eq. (11) is rewritten as
\begin{equation*}
Q^{(N)}=\left(
\begin{array}{ccc}
0&0&u-2\text{i}\Theta_{13}\\
0&0&\epsilon u^{*}-2\text{i}\Theta_{23}\\
\epsilon(au^{*}+bu)+2\text{i}\Theta_{31}&au+bu^{*}+2\text{i}\Theta_{32}&0
\end{array}
\right),
\end{equation*}
where
\begin{equation}\begin{aligned}
&\Theta_{13}=-\frac{\left|
\begin{array}{cc}
W_N&\mathbf{h}_3^{\dag}\\ \mathbf{h}_1&0
\end{array}
\right|}{|W_N|},~~~~\Theta_{31}=\frac{\epsilon a\left|
\begin{array}{cc}
W_N&\mathbf{h}_1^{\dag}\\ \mathbf{h}_3&0
\end{array}
\right|+b\left|
\begin{array}{cc}
W_N&\mathbf{h}_2^{\dag}\\ \mathbf{h}_3&0
\end{array}
\right|}{|W_N|},\\ \nonumber \\ \nonumber
&\Theta_{23}=-\frac{\left|
\begin{array}{cc}
W_N&\mathbf{h}_3^{\dag}\\ \mathbf{h}_2&0
\end{array}
\right|}{|W_N|},~~~~\Theta_{32}=\frac{b\left|
\begin{array}{cc}
W_N&\mathbf{h}_1^{\dag}\\ \mathbf{h}_3&0
\end{array}
\right|+\epsilon a\left|
\begin{array}{cc}
W_N&\mathbf{h}_2^{\dag}\\ \mathbf{h}_3&0
\end{array}
\right|}{|W_N|},
\end{aligned}\end{equation}
with
\begin{equation*}
\mathbf{h}_2=(\psi_{2}^{(1)},\psi_{1}^{(1)^*},\psi_{2}^{(2)},\psi_{1}^{(2)^*},\cdots,\psi_{2}^{(N)},\psi_{1}^{(N)^*}).
\end{equation*}
Note that
\begin{equation*}
\langle \eta_{i}|J|\eta_{k}\rangle=\langle y_{k}|J|y_{i}\rangle=(\langle y_{i}|J|y_{k}\rangle)^*,~~\langle \eta_{i}|J|y_{k}\rangle=(\langle y_{i}|J|\eta_{k}\rangle)^*,~~\langle y_{k}|J|\eta_{i}\rangle=\langle y_{i}|J|\eta_{k}\rangle,
\end{equation*}
we have $\Omega(K_i,K_k)+\Omega(K_k,K_i)^{\dag}=0$, and $W_N$ is skew Hermitian matrix, i.e.$W_N+W_N^{\dag}=0$. By introducing $2N\times2N$ permutation matrix
\begin{equation*}
A=\left(
\begin{array}{ccccc}
0&1&\cdots&0&0\\
1&0&\cdots&0&0\\
\vdots&\vdots&\ddots&\vdots&\vdots\\
0&0&\cdots&0&1\\
0&0&\cdots&1&0
\end{array}
\right),
\end{equation*}
we have $\mathbf{h}_1A=\mathbf{h}_2^*,\mathbf{h}_2A=\mathbf{h}_1^*,\mathbf{h}_3A=\epsilon \mathbf{h}_3^*$ and $AW_NA=W_N^{T}$. With these identities, we have
\begin{equation}\begin{aligned}
&\left|
\begin{array}{cc}
W_N&\mathbf{h}_3^{\dag}\\ \mathbf{h}_1&0
\end{array}
\right|=\left|
\begin{array}{cc}
AW_N^{T}A&\epsilon A\mathbf{h}_3^{\dag}\\ \mathbf{h}_2^*A&0
\end{array}
\right|=\left|
\begin{array}{cc}
W_N^T&\epsilon\mathbf{h}_3^{T}\\ \mathbf{h}_2^*&0
\end{array}
\right|=\left|
\begin{array}{cc}
W_N&\mathbf{h}_2^{\dag}\\ \epsilon\mathbf{h}_3&0
\end{array}
\right|\\ \nonumber\\ \nonumber
&\qquad\qquad~~~~=-\left|
\begin{array}{cc}
W_N^{\dag}&\mathbf{h}_2^{\dag}\\ \epsilon\mathbf{h}_3&0
\end{array}
\right|=-\epsilon\left|
\begin{array}{cc}
W_N^{*}&\mathbf{h}_3^{T}\\ \mathbf{h}_2^*&0
\end{array}
\right|=-\epsilon\left|
\begin{array}{cc}
W_N&\mathbf{h}_3^{\dag}\\ \mathbf{h}_2&0
\end{array}
\right|^*,
\end{aligned}\end{equation}
and
\begin{equation}
\Theta_{23}=-\epsilon\Theta_{13}^*,~~\Theta_{31}=\epsilon(a\Theta_{13}^*-b\Theta_{13}),
~~\Theta_{32}=-a\Theta_{13}+b\Theta_{13}^*.
\end{equation}
This completes the proof of our main result.


\section{ \bf Hump-soliton and breather solutions for Eq. (4) with the zero seed solution}

In the section, by using the DT, we obtain hump-soliton solution, breather solution and hump-breather solution for Eq. (4) with the zero seed solution. The asymptotic behavior of $2$-soliton and $2$-breather solutions is analyzed.

For the zero seed solution $u=0$, solving the eigenvalue problem (6) with eigenvalue $\lambda_k=\alpha_k+\text{i}\beta_k$$(\beta_k\neq0)$, yields the eigenfunction
\begin{equation}\label{Y0}
\psi_{1}^{(k)}=c_{3k-2}\text{e}^{\theta_{k}},\psi_{2}^{(k)}=c_{3k-1}\text{e}^{\theta_{k}},\psi_{3}^{(k)}=c_{3k}\text{e}^{-\theta_{k}},\theta_{k}=\text{i}\lambda_k(x+4\lambda_k^2t), k=1,2,...,N,
\end{equation}
where $c_{j}$ ($j=1,2,\cdots,3N$) are complex constants.

\subsection{ $1$-Hump soliton and breather solutions }

By using $1$-fold DT, we obtain the soliton solution for the gSS equation (4)
\begin{equation}\label{solution2}
u^{(1)}=-\frac{2\text{i}}{A_1^2+|B_1|^2}\left(A_1(\psi_{1}^{(1)}\psi_{3}^{(1)^*}+\epsilon\psi_{2}^{(1)^*}\psi_{3}^{(1)})-\epsilon B_1\psi_{1}^{(1)}\psi_{3}^{(1)}+B_1^*\psi_{2}^{(1)^*}\psi_{3}^{(1)^*}\right)
\end{equation}
where
\begin{equation}\label{1-DT}
A_1=\frac{\langle y_{1}|J|y_{1}\rangle}{\lambda_1^*-\lambda_1},~~B_1=\frac{\langle y_{1}|J|\eta_{1}\rangle}{2\lambda_1^*},~~|y_{1}\rangle=(\psi_{1}^{(1)},\psi_{2}^{(1)},\psi_{3}^{(1)})^T,~~ |\eta_{1}\rangle=(\psi_{2}^{(1)^*},\psi_{1}^{(1)^*},\epsilon\psi_{3}^{(1)^*})^{T}.
\end{equation}
and $c_3\neq0$, $|c_1|^2+|c_2|^2\neq0$. Set $c_1=1$, $c_3=1$, $\lambda_1=\alpha_1+\text{i}\beta_1$. The solution can be written as
\begin{equation}\label{u1}
u^{(1)}=\frac{4\beta_1(\alpha_1(\epsilon c_2^*\lambda_1^*\text{e}^{-2\theta_{1}}+\lambda_1\text{e}^{-2\theta_{1}^*})-\lambda_1^*(\text{i}\nu_1+\alpha_1\omega_1)\text{e}^{2\theta_{1}}-\epsilon\lambda_1(\text{i}\nu_2+\alpha_1c_2^*\omega_1)\text{e}^{2\theta_{1}^*})}
{\nu_3-2\alpha_1^2\text{cosh}[4\theta_{1,R}]-2\beta_1^2\omega_3\text{cos}[4\theta_{1,I}]+4\text{i}c_{2,I}\beta_1^2(a+\epsilon bc_{2,R})\text{e}^{-4\text{i}\theta_{1,I}}},
\end{equation}
where superscript $R$, $I$ represent real part and imaginary part, and
\begin{eqnarray*}
&&\omega_1=a\epsilon(1+|c_2|^2)+2bc_{2,R},~\omega_2=\beta_1^2(b^2-a^2)(1-|c_2|^2)^2,~\omega_3=2ac_2+\epsilon b(1+c_2^2),\\
&&\nu_1=\beta_1(a\epsilon+bc_2)(1-|c_2|^2),~\nu_2=\beta_1(b+a\epsilon c_2^*)(1-|c_2|^2),\\
&&\nu_3=2|\lambda_1|^2\omega_1+(\alpha_1^2(1-\omega_1^2)+\omega_2)\text{e}^{4\theta_{1,R}}.
\end{eqnarray*}
It can be seen that the solution (25) is singular for the defocusing gSS equation (4). Let us discuss two cases for the focusing gSS equation (4).

\textbf{Case 1.} Set $c_2=0$ and $a=-1$. The solution of the gSS equation (4) is given by
\begin{equation}\label{fig0}
u^{(1)}=-\frac{4\beta_1(\alpha_1+\text{i}\beta_1)\text{e}^{2\text{i}\theta_{1,I}}(2\alpha_1\text{cosh}(2\theta_{1,R})
-\text{i}\beta_1\text{e}^{2\theta_{1,R}}-\text{i}\beta_1b\text{e}^{2\theta_{1,R}-4\text{i}\theta_{1,I}})}
{2(\alpha_1^2+\beta_1^2)+2\alpha_1^2\text{cosh}(4\theta_{1,R})+\beta_1^2(1-b^2)\text{e}^{4\theta_{1,R}}+2\beta_1^2b\text{cosh}(4\theta_{1,I})}.
\end{equation}
If set $b=0$, we obtain the hump-soliton solution of the Sasa-Satsuma equation (3). We should remark that the solution (26) is different from the solution obtained in [5].
When $\beta_1^2\leq3\alpha_1^2$, this solution is single-hump soliton, and the module $|u^{(1)}|$ reaches to its maximum at the line
\begin{equation*}
x=4(\beta_1^2-3\alpha_1^2)t+\frac{1}{4\beta_1}\ln(\frac{\sqrt{\alpha_1^2+\beta_1^2}}{|\alpha_1|});
\end{equation*}
when $\beta_1^2>3\alpha_1^2$, this solution is double-hump soliton, and the module $|u^{(1)}|$ reaches to its maximum at lines
\begin{equation*}
x=4(\beta_1^2-3\alpha_1^2)t+\frac{1}{4\beta_1}\ln(\frac{\beta_1^2\pm\sqrt{\beta_1^2(\beta_1^2-3\alpha_1^2)}}{\alpha_1^2}-1).
\end{equation*}

Fig.1 shows the process of from single-hump soliton to double-hump soliton for $\beta_1^2<3\alpha_1^2$,$\beta_1^2=3\alpha_1^2$ and $\beta_1^2>3\alpha_1^2$, respectively.

Our focus is the gSS equation for $b\neq0$(e.g. $b=\frac{1}{2}$). The soliton solution of the gSS equation is given by
\begin{eqnarray}\label{fig02}
&u^{(1)}=-\frac{8\beta_1(\alpha_1+\text{i}\beta_1)\text{e}^{2\text{i}\theta_{1,I}}(4\alpha_1\text{cosh}(2\theta_{1,R})
-2\text{i}\beta_1\text{e}^{2\theta_{1,R}}-\text{i}\beta_1\text{e}^{2\theta_{1,R}-4\text{i}\theta_{1,I}})}
{8(\alpha_1^2+\beta_1^2)+8\alpha_1^2\text{cosh}(4\theta_{1,R})+3\beta_1^2\text{e}^{4\theta_{1,R}}+4\beta_1^2\text{cosh}(4\theta_{1,I})}.
\end{eqnarray}
We can see that the solution displays a breather-like form traveling along the peak line of the case of $b=0$. When $\beta_1^2<3\alpha_1^2$, this solution is single-peak breather-like; when $\beta_1^2=3\alpha_1^2$, this solution is Kuznetsov-Ma(KM) breather-like; When $\beta_1^2>3\alpha_1^2$, this solution is double-peak breather-like. Fig.2 shows that the process of changing single-peak breather-like solution to double-peak breather-like solution. We should emphasize here that there exist a big difference between the SS equation and the gSS equation.

\textbf{Case 2:} $c_2\neq0$(e.g. $c_2=1$). We obtain the breather solution of the gSS equation
\begin{eqnarray}\label{fig03}
&u^{(1)}=\frac{8\alpha_1\beta_1\left(2\text{cosh}(2\theta_{1,R})(\alpha_1\text{cos}(2\theta_{1,I})-\beta_1\text{sin}(2\theta_{1,I}))
-\text{e}^{2\theta_{1,R}}(\alpha_1\varrho_1\text{cos}(2\theta_{1,I})-\beta_1\varrho_2\text{sin}(2\theta_{1,I}))\right)}
{4(a+b)(\alpha_1^2+\beta_1^2)-2\alpha_1^2\text{cosh}(4\theta_{1,R})-4(a+b)\beta_1^2\text{cos}(4\theta_{1,R})+\alpha_1^2\varrho_1\varrho_2\text{e}^{4\theta_{1,R}}},
\end{eqnarray}
where $\varrho_1=1+2(a+b)$ and $\varrho_2=1-2(a+b)$. When $\beta_1^2=3\alpha_1^2$, this solution is KM-breather solution. When $\beta_1^2\neq3\alpha_1^2$, this solution is a general breather solution in space and time. In Fig.3, we give plots of such breather solutions for the gSS equation(4).

\subsection{ $2$-soliton and breather solutions }

For the focusing gSS equation (4), by using $2$-fold DT, we obtain $2$-soliton solution
\begin{equation}\label{solution3}
u^{(2)}=-2\text{i}(\psi_1^{(1)},\psi_2^{(1)^*},\psi_1^{(2)},\psi_2^{(2)^*})
W_2^{-1}(\psi_3^{(1)^*},\psi_3^{(1)},\psi_3^{(2)^*},\psi_3^{(2)})^T
\end{equation}
where
\begin{eqnarray*}
&&W_2^{-1}=\frac{1}{G}\left(
\begin{array}{cccc}
R_1&T_1&T_2&T_3\\
-T_1^*&R_1&-T_3^*&-T_2^*\\
-T_2^*&T_3&R_2&T_4\\
-T_3^*&T_2&-T_4^*&R_2
\end{array}
\right),\\
&&G=\left(A_1A_2+|C_1|^2+|D_1|^2\right)^2+A_1^2|B_2|^2+|B_1|^2(A_2^2+|B_2|^2)-4|C_1|^2|D_1|^2\\
&&~~~~+2Re\left[B_1(C_1^{*^2}B_2^*-D_1^{*^2}B_2)+2D_1^*(A_2B_1C_1^*-A_1C_1B_2)\right],\\
&&R_1=C_1(A_2C_1^*-D_1^*B_2)+D_1(A_2D_1^*+C_1^*B_2^*)+A_1(A_2^2+|B_2|^2),\\
&&R_2=A_2(A_1^2+|B_1|^2)+C_1(A_1C_1^*-B_1^*D_1)+D_1^*(B_1C_1^*+A_1D_1),\\
&&T_1=2A_2C_1D_1-A_2^2B_1-C_1^2B_2+B_2^*(D_1^2-B_1B_2),\\
&&T_2=C_1(|D_1|^2-|C_1|^2)-B_1(A_2D_1^*+C_1^*B_2^*)-A_1(A_2C_1+D_1B_2^*),\\
&&T_3=C_1(C_1^*D_1+A_1B_2)-A_2(B_1C_1^*+A_1D_1)-D_1^*(D_1^2-B_1B_2),\\
&&T_4=B_1^*D_1^2-2A_1C_1^*D_1-A_1^2B_2-B_1(C_1^{*^2}+B_1^*B_2),\\
&&A_k=\frac{\langle y_k|J|y_k\rangle}{\lambda_k^*-\lambda_k},~B_k=\frac{\langle y_k|J|\eta_k\rangle}{2\lambda_k^*},k=1,2,~C_1=\frac{\langle y_1|J|y_2\rangle}{\lambda_1^*-\lambda_2},~D_1=\frac{\langle y_1|J|\eta_2\rangle}{\lambda_1^*+\lambda_2^*}.
\end{eqnarray*}
We remark here that the matrix $W_2^{-1}$ is complex, but we need it when we analyze the asymptotic behavior of the $2$-soliton. Let us discuss the property for this solution. We first consider the case of the focusing SS equation(3).

\textbf{Case 1.} Set $c_1=c_3=c_5=c_6=1$, $c_2=c_4=0$.

If $\lambda_{1,R}\lambda_{1,I}\lambda_{2,R}\lambda_{2,I}\neq 0$, we can obtain the asymptotic behavior for the interaction of this solution.

When $\theta_1\sim O(1)$, we have
\begin{eqnarray*}
\text{if}~~ \lambda_{2,I}(3\lambda_{1,R}^2-\lambda_{1,I}^2-3\lambda_{2,R}^2+\lambda_{2,I}^2)>0,~u^{(2)}\sim \begin{cases}
u_{1}^{-},\quad t\rightarrow-\infty,\\
u_{1}^{+},\quad t\rightarrow+\infty,
\end{cases}\\
\text{if}~~ \lambda_{2,I}(3\lambda_{1,R}^2-\lambda_{1,I}^2-3\lambda_{2,R}^2+\lambda_{2,I}^2)<0,~u^{(2)}\sim \begin{cases}
u_{1}^{+},\quad t\rightarrow-\infty,\\
u_{1}^{-},\quad t\rightarrow+\infty,
\end{cases}
\end{eqnarray*}
where
\begin{equation*}
u_{1}^{-}=\frac{\mathbf{u}_1}{\mathbf{u}_2},
u_{1}^{+}=\frac{\mathbf{u}_3}{\mathbf{u}_4},
\end{equation*}
with
\begin{eqnarray*}
&\mathbf{u}_1=4\lambda_1\lambda_{1,I}(\lambda_1+\lambda_2)(\lambda_2^*-\lambda_1)(\lambda_1^*+\lambda_2)(\lambda_1^*-\lambda_2^*)S_1,\\
&\mathbf{u}_2=2|\lambda_1|^2|\lambda_1^2-\lambda_2^{*2}|^2|\lambda_1^2
-\lambda_2^2|^2+|\lambda_1|^2|\lambda_1+\lambda_2|^4|\lambda_1-\lambda_2^*|^4\text{e}^{4\theta_{1,R}}
\\
&~~~~~~+\lambda_{1,R}^2|\lambda_1-\lambda_2|^4|\lambda_1+\lambda_2^*|^4\text{e}^{-4\theta_{1,R}},\\
&\mathbf{u}_3=4\lambda_1\lambda_{1,I}(\lambda_1-\lambda_2)(\lambda_2-\lambda_1^*)
(\lambda_{1,R}|\lambda_1-\lambda_2^*|^2\text{e}^{-2\theta_1^*}+\lambda_1^*|\lambda_1-\lambda_2|^2\text{e}^{2\theta_1}),\\
&\mathbf{u}_4=2|\lambda_1|^2|\lambda_1-\lambda_2|^2|\lambda_1-\lambda_2^{*}|^2+|\lambda_1|^2|\lambda_1-\lambda_2|^4\text{e}^{4\theta_{1,R}}
+\lambda_{1,R}^2|\lambda_1-\lambda_2|^4\text{e}^{-4\theta_{1,R}},\\
&S_1=\lambda_{1,R}|\lambda_1-\lambda_2|^2|\lambda_1+\lambda_2^*|^2\text{e}^{-2\theta_1^*}+\lambda_1^*|\lambda_1+\lambda_2|^2|\lambda_1-\lambda_2^*|^2\text{e}^{2\theta_1}.
\end{eqnarray*}

When $\theta_2\sim O(1)$,
\begin{eqnarray*}
\text{if}~~ -\lambda_{1,I}(3\lambda_{1,R}^2-\lambda_{1,I}^2-3\lambda_{2,R}^2+\lambda_{2,I}^2)>0,~u^{(2)}\sim \begin{cases}
u_{2}^{-},\quad t\rightarrow-\infty,\\
u_{2}^{+},\quad t\rightarrow+\infty,
\end{cases}\\
\text{if}~~ -\lambda_{1,I}(3\lambda_{1,R}^2-\lambda_{1,I}^2-3\lambda_{2,R}^2+\lambda_{2,I}^2)<0,~u^{(2)}\sim \begin{cases}
u_{2}^{+},\quad t\rightarrow-\infty,\\
u_{2}^{-},\quad t\rightarrow+\infty,
\end{cases}
\end{eqnarray*}
with
\begin{eqnarray*}
&u_{2}^{-}=\frac{4\lambda_2\lambda_{2,I}(\lambda_1+\lambda_2)(\lambda_1^*-\lambda_2)(\lambda_2^*-\lambda_1^*)(\lambda_1+\lambda_2^*)S_2}
{2|\lambda_2|^2|\lambda_1^2-\lambda_2^{*2}|^2|\lambda_1^2-\lambda_2^2|^2+|\lambda_2|^2|\lambda_1+\lambda_2|^4|\lambda_1-\lambda_2^*|^4\text{e}^{4\theta_{1,R}}
+\lambda_{2,R}^2|\lambda_1-\lambda_2|^4|\lambda_1+\lambda_2^*|^4\text{e}^{-4\theta_{1,R}}},\\
&u_{2}^{+}=\frac{4\lambda_2\lambda_{2,I}(\lambda_1-\lambda_2)(\lambda_2^*-\lambda_1)
(\lambda_{2,R}|\lambda_1-\lambda_2^*|^2\text{e}^{-2\theta_2^*}+\lambda_2^*|\lambda_1-\lambda_2|^2\text{e}^{2\theta_2})}
{2|\lambda_2|^2|\lambda_1-\lambda_2|^2|\lambda_1-\lambda_2^*|^2+|\lambda_2|^2|\lambda_1-\lambda_2|^4\text{e}^{4\theta_{2,R}}
+\lambda_{2,R}^2|\lambda_1-\lambda_2^*|^4\text{e}^{-4\theta_{2,R}}},
\end{eqnarray*}
with
\begin{equation*}
S_2=\lambda_{2,R}|\lambda_1-\lambda_2|^2|\lambda_1+\lambda_2^*|^2\text{e}^{-2\theta_2^*}+\lambda_2^*|\lambda_1+\lambda_2|^2|\lambda_1-\lambda_2^*|^2\text{e}^{2\theta_1}.
\end{equation*}

Setting $\lambda_1=1+\frac{\text{i}}{2}$,$\lambda_2=\frac{1}{3}+\text{i}$, the $2$-soliton solution for the Sasa-Satsuma equation(3) describes the elastic collision of single-hump soliton with double-hump soliton(see Fig. 4$(a)$). With above analysis, the asymptotic behavior for this solution can be written as
\begin{equation}\label{asymptotic0}
u^{(2)}\longrightarrow \begin{cases}
u_{1}^{-}+u_{2}^{+},\quad t\rightarrow-\infty,\\
u_{1}^{+}+u_{2}^{-},\quad t\rightarrow+\infty,
\end{cases}
\end{equation}
where
\begin{eqnarray*}
&u_{1}^{-}=\frac{(2006-288\text{i})(2813\text{e}^{2\theta_1}+146(2+\text{i})\text{e}^{-2\theta_1^{*}})}
{730(292\text{cosh}(4\theta_{1,R})+2813)+7806389\text{e}^{4\theta_{1,R}}},~~~
u_{2}^{+}=\frac{4(29-153\text{i})(25(1-3\text{i})\text{e}^{2\theta_2}+97\text{e}^{-2\theta_2^{*}})}
{500(25\text{cosh}(4\theta_{2,R})+97)+3159\text{e}^{-4\theta_{2,R}}},\\
&u_{1}^{+}=\frac{(214\text{i}-52)(25(2-\text{i})\text{e}^{2\theta_1}+194\text{e}^{-2\theta_1^*})}
{250(25\text{cosh}(4\theta_{1,R})+97)+34511\text{e}^{-4\theta_{1,R}}},~~
u_{2}^{-}=\frac{4(1037+989\text{i})(2813(\text{i}-1)\text{e}^{2\theta_2}-73(2+\text{i})\text{e}^{-2\theta_2^*})}
{730(73\text{cosh}(4\theta_{2,R})+5626)+15799293\text{e}^{4\theta_{2,R}}}.
\end{eqnarray*}
It can be checked that $u_{1}^{-},u_{1}^{+}$ are single-hump soliton and $u_{2}^{-},u_{2}^{+}$ are double-hump soliton solutions of the focusing Sasa-Satsuma equation(3).
The peak values of $u_{1}^{-}$ and $u_{1}^{+}$ are same, which take at the lines
\begin{equation*}
x=-11t+\frac{1}{2}\ln(\frac{2813}{146\sqrt{5}})~~ \text{and}~~ x=-11t+\frac{1}{2}\ln(\frac{25\sqrt{5}}{194}),
\end{equation*}
respectively. The peak values of $u_{2}^{-}$ and $u_{2}^{+}$ are same, which take at the lines
\begin{equation*}
x=\frac{1}{3}\left(8t-\ln(\frac{97^{\frac{3}{4}}}{5\times2^{\frac{3}{8}}\times5^{\frac{7}{8}}})\pm\ln(\frac{\sqrt[4]{904+369\sqrt{6}}}{2^{1/8}\times5^{3/8}})\right),
\end{equation*}
and
\begin{equation*}
x=\frac{1}{3}\left(8t-\ln((\frac{73}{2813})^{\frac{3}{4}}(\frac{5}{2})^{\frac{3}{8}})\pm\ln(\frac{\sqrt[4]{904+369\sqrt{6}}}{2^{1/8}\times5^{3/8}})\right),
\end{equation*}
respectively. It can be seen that the interaction between single-hump soliton and double-hump soliton is elastic.

If take $\lambda_1=1+\frac{\text{i}}{2},\lambda_2=\frac{3}{2}+2\text{i}$, ($\lambda_{2,I}(3\lambda_{1,R}^3-\lambda_{1,I}-3\lambda_{2,R}^3+\lambda_{2,I})=0$), the solution of the focusing Sasa-Satsuma equation(3) is double-peak breather(see Fig. 4$(b)$); If take $\lambda_1=1+\frac{\text{i}}{2},\lambda_2=\text{i}$, this solution of the focusing Sasa-Satsuma equation(3) displays the interaction of single-peak breather and hump soliton, one breather becomes into single-hump soliton after the collision(see Fig. 4$(c)$);

Let us give a analysis for the interaction of two solitons to the focusing gSS equation$(4)$ with $a=-1,b=\frac{1}{2}$. Set $\lambda_1=1+\frac{\text{i}}{2}$,$\lambda_2=\frac{1}{3}+\text{i}$, the $2$-soliton solution describes the collision between single-peak breather and double-peak breather soliton(see Fig. 4$(d)$). The asymptotic behavior for this $2$-soliton solution is analyzed as follows
\begin{equation}\label{asymptotic1}
u^{(2)}\longrightarrow \begin{cases}
u_{12}^{-}+u_{22}^{-},\quad t\rightarrow-\infty,\\
u_{12}^{+}+u_{22}^{+},\quad t\rightarrow+\infty,
\end{cases}
\end{equation}
where
\begin{eqnarray*}
&u_{12}^{-}=\frac{4(49-108\text{i})\left((52871+9638\text{i})\text{e}^{2\theta_1^*}+5626(23+36\text{i})\text{e}^{2\theta_1}+1460(2+19\text{i})\text{e}^{-2\theta_1^*}\right)}
{4(288864\text{sin}(4\theta_{1,I})+985273\text{cos}(4\theta_{1,I}))+14600(2813+292\text{cosh}(4\theta_{1,R}))+148214811\text{e}^{4\theta_{1,R}}},\\
&u_{22}^{-}=\frac{45008A_1-200A_2}
{36B_1+50(55517368+12446221\text{cosh}(4\theta_{2,R}))+290230119\text{e}^{-4\theta_{2,R}}},\\
&u_{12}^{+}=\frac{100A_3+45008A_4}
{4B_2+50(49767596+12446221\text{cosh}(4\theta_{1,R}))+3613677099\text{e}^{-4\theta_{1,R}}},\\
&u_{22}^{+}=\frac{-8(149+77\text{i})(3(38393-37606\text{i})\text{e}^{2\theta_2^*}+5626(53-29\text{i})\text{e}^{2\theta_2})+730(14+13\text{i})\text{e}^{-2\theta_2^*}}
{36(1025593\text{cos}(4\theta_{2,I})+48624\text{sin}(4\theta_{2,I})+14600(5626+73\text{cosh}(4\theta_{2,R}))+244769139\text{e}^{4\theta_{2,R}}},\\
&A_1=(13231-87867\text{i})\text{e}^{-2\theta_2^*}+120(73+151\text{i})\text{e}^{-2\theta_2},\\
&A_2=3(1604087-652757\text{i})\text{e}^{2\theta_2^*}+580(23291+11010\text{i})\text{e}^{2\theta_2},\\
&A_3=290(12469+44040\text{i})\text{e}^{2\theta_1}+(160085-1913998\text{i})\text{e}^{2\theta_1^*},\\
&A_4=120(47-161\text{i})\text{e}^{-2\theta_1}-(17938-104041\text{i})\text{e}^{-2\theta_1^*},\\
&B_1=12373027\text{cos}(4\theta_{2,I})-100536\text{sin}(4\theta_{2,I}),\\
&B_2=4941576\text{sin}(4\theta_{1,I})-32387057\text{cos}(4\theta_{1,I}).
\end{eqnarray*}
It can be proved that $u_{21}^{-},u_{21}^{+}$ are single-peak breather solution and $u_{22}^{-},u_{22}^{+}$ are double-peak breather solution of Eq.$(4)$. If take $\lambda_1=1+\frac{\text{i}}{2},\lambda_2=\frac{3}{2}+2\text{i}$, this solution is periodic-like solution(see Fig. 4$(e)$); If take $\lambda_1=1+\frac{\text{i}}{2},\lambda_2=\text{i}$, this solution displays the collision of breather and hump soliton, the breather soliton becomes into breather-like soliton after the collision(see Fig.4$(f)$).

\textbf{Case 2.} Set $c_2=0,c_4=1$, $c_1=c_3=c_5=c_6=1$, and $\lambda_1=1+\frac{\text{i}}{2}$, $\lambda_2=\frac{1}{3}+\text{i}$, $a=-1$.

For the focusing Sasa-Satsuma equation(3), the $2$-soliton solution describes that a single-hump soliton changes into single-peak breather after colliding with a breather soliton(see Fig. 5$(a)$). The asymptotic behavior for this solution is analyzed as follows
\begin{equation}\label{asymptotic2}
u^{(2)}\longrightarrow \begin{cases}
u_{13}^{-}+u_{23}^{-},\quad t\rightarrow-\infty,\\
u_{13}^{+}+u_{23}^{+},\quad t\rightarrow+\infty;
\end{cases}
\end{equation}
where $u_{13}^{-}=u_{11}^{-}$ and
\begin{eqnarray*}
&u_{23}^{-}=\frac{8((784742+94796\text{i})B_1\text{e}^{2\theta_{2,R}}+14065B_2\text{e}^{-2\theta_{2,R}})}
{506340(1013\text{cos}(4\theta_{2,I})-24\text{sin}(4\theta_{2,I}))+53(91659\text{e}^{-4\theta_{2,R}}-11364520)-88845544\text{cosh}(4\theta_{2,R})},\\
&u_{13}^{+}=\frac{2((1003+3178\text{i})B_3+5626B_4)}
{144(48109\text{cos}(4\theta_{1,I})+2006\text{sin}(4\theta_{1,I}))+5626(7945+11252\text{cosh}(4\theta_{1,R}))-20546183\text{e}^{4\theta_{1,R}}},\\
&u_{23}^{+}=\frac{8(5626\text{e}^{2\theta_{2,R}}(941\text{cos}(2\theta_{2,I})+3063\text{sin}(2\theta_{2,I}))+1825(217\text{cos}(2\theta_{2,I})-603\text{sin}(2\theta_{2,I})))}
{36(1025593\text{cos}(4\theta_{2,I})+48624\text{sin}(4\theta_{2,I}))-3650(73\text{cosh}(4\theta_{2,R})+11252)-31518651\text{e}^{-2\theta_{2,R}}},\\
&B_1=(37+2\text{i})\text{cos}(2\theta_{2,I})+(87-42\text{i})\text{sin}(2\theta_{2,I}),\\
&B_2=(1859+36\text{i})\text{cos}(2\theta_{2,I})-27(163+100\text{i})\text{sin}(2\theta_{2,I}),\\
&B_3=36(49-108\text{i})\text{e}^{2\text{i}\theta_1^*}+(1049+1738\text{i})\text{e}^{2\text{i}\theta_1},\\
&B_4=36(104-57\text{i})\text{e}^{-2\text{i}\theta_1}-(1882+761\text{i})\text{e}^{-2\text{i}\theta_1^*}.
\end{eqnarray*}
It can be proved that $u_{13}^{+},u_{23}^{-}$ are single-peak breather solutions of the focusing Sasa-Satsuma equation, and $u_{23}^{+}$ is a breather solution of the corresponding focusing mKdV equation.

Let us analyze the case of the gSS equation. If $b\neq0$ (e.g.$b=\frac{1}{2}$), the $2$-soliton solution shows that a breather-like wave changes into breather wave after colliding with a breather wave(see Fig. 5$(b)$). The asymptotic behavior for this solution is analyzed as follows
\begin{equation}\label{asymptotic11}
u^{(2)}\longrightarrow \begin{cases}
u_{14}^{-}+u_{24}^{-},\quad t\rightarrow-\infty,\\
u_{14}^{+}+u_{24}^{+},\quad t\rightarrow+\infty,
\end{cases}
\end{equation}
where $u_{14}^{-}=u_{12}^{-}$ and
\begin{eqnarray*}
&u_{24}^{-}=\frac{4(2-\text{i})(C_1\text{e}^{2\theta_{2,R}}+2813C_2\text{e}^{-2\theta_{2,R}})}
{9(88170163\text{cos}(4\theta_{2,I})-915432\text{sin}(4\theta_{2,I}))-2813(322070+53447\text{cosh}(4\theta_{2,R}))+34908984\text{e}^{4\theta_{2,R}}},\\
&u_{14}^{+}=\frac{11252C_3-4C_4}{4(11376817\text{cos}(4\theta_{1,I})+144432\text{sin}(4\theta_{1,I})+61914130)+161056886\text{cosh}(4\theta_{1,R})+46079061\text{e}^{-4\theta_{1,R}}},\\
&u_{24}^{+}=\frac{4(2813\text{e}^{2\theta_{2,R}}(941\text{cos}(2\theta_{1,I})+3063\text{sin}(2\theta_{1,I}))+1825\text{e}^{-2\theta_{2,R}}(217\text{cos}(2\theta_{1,I})-603\text{sin}(2\theta_{1,I})))}
{9(1025593\text{cos}(4\theta_{2,I})+48624\text{sin}(4\theta_{2,I}))-1825(5626+73\text{cosh}(4\theta_{2,R}))-3889872\text{e}^{4\theta_{2,R}}},
\end{eqnarray*}
with \begin{eqnarray*}
&C_1=(34319674+21315209\text{i})\text{cos}(2\theta_{1,I})+3(35203162+12464833\text{i})\text{sin}(2\theta_{1,I}),\\
&C_2=(16706+9469\text{i})\text{cos}(2\theta_{1,I})-(37734+35055\text{i})\text{sin}(2\theta_{1,I}),\\
&C_3=36(104-57\text{i})\text{e}^{-2\theta_1}-(1882+761\text{i})\text{e}^{-2\theta_1^*},\\
&C_4=2(5713883-8310744\text{i})\text{e}^{2\theta_1}-(38820683+5987714\text{i})\text{e}^{2\theta_1^*}.
\end{eqnarray*}
It can be proved that $u_{14}^{+}$, $u_{24}^{-}$ are breather solutions of focusing gSS equation$(4)$ with $a=-1,b=\frac{1}{2}$, and $u_{24}^{+}$ is a breather solution of the corresponding focusing mKdV equation.

\section{Breather and periodic solutions for Eq.(4) with the nonzero seed solution}

Starting from the nonzero seed solution, we present the bright-dark breather soliton, bright-bright breather soliton, resonant$(2,1)$ interaction(i.e the solution shows that two soliton become a soliton after resonance) and general periodic solutions for the focusing gSS equation. Meanwhile, we obtain the dark breather solution of the defocusing gSS equation. Note that bright-dark$(i,j)$(i.e. the number of peaks and troughs of the breather in shape are $i,j$ respectively), bright-bright breather and resonant$(2,1)$-breather interaction solution of the Sasa-Satsuma equation have not been presented in the literatures.

For the nonzero seed solution $u=\gamma$($\gamma\neq0$,$\gamma$ is a real constant), solving the eigenvalue problem (4) at $\lambda=\lambda_j$, gives the the eigenfunction
\begin{equation}\label{eigenfunction}\begin{aligned}
&\Psi_j=(\psi_1^{(j)},\psi_2^{(j)},\psi_3^{(j)})^T,~~\psi_1^{(j)}=\gamma(d_{j1}\text{e}^{\chi_j}+d_{j2}\text{e}^{-\chi_j}+d_{j3}\text{e}^{\xi_j}),\\ \nonumber
&\psi_2^{(j)}=\epsilon\gamma(d_{j1}\text{e}^{\chi_j}+d_{j2}\text{e}^{-\chi_j}-d_{j3}\text{e}^{\xi_j}),~~\psi_3^{(j)}=d_{j1}\eta_j\text{e}^{\chi_j}-d_{j2}\kappa_j\text{e}^{-\chi_j},\\ \nonumber
&\chi_j=\tau_j(x+4\delta_1t),\xi_j=\text{i}\lambda_j(x+4\lambda_j^2t),\eta_j=-\text{i}\lambda_j+\tau_j,\kappa_j=\text{i}\lambda_j+\tau_j,\nonumber
\end{aligned}\end{equation}
where $d_{j1},d_{j2},d_{j3}$ are complex constants and $\tau_j=\sqrt{2\epsilon(a+b)\gamma^2-\lambda_j^2}$,
$\delta_j=\epsilon(a+b)\gamma^2+\lambda_j^2$.

\subsection{ $1$-soliton and breather solutions }

Using $1$-fold DT yields soliton solution of the gSS equation. Here we set parameters $\gamma=1$.

If $d_{11}=1,d_{12}=0,d_{13}=1$, the solution can be written as
\begin{equation}\label{solution3-2}\begin{aligned}
&u^{(1)}=1+\frac{16\epsilon\lambda_{1,I}(4\omega_1H_1+\text{i}\text{Im}[p_1\text{e}^{\chi_1^*-\xi_1^*}+4\lambda_{1,R}\omega_1\lambda_1^*\eta_1^*\text{e}^{\xi_1-\chi_1}]+\lambda_{1,R}p_2\text{e}^{2(\chi_{1,R}-\xi_{1,R})}}
{16\omega_1(H_2+\lambda_{1,I}^2\text{Re}[(2\omega_2-\epsilon\eta_1^2)\text{e}^{2\text{i}(\chi_{1,I}-\xi_{1,I})}])+p_3\text{e}^{2(\chi_{1,R}-\xi_{1,R})}
},\\
&\omega_1=a-b,~~\omega_2=a+b,~~p_1=2\lambda_1\omega_2(2\lambda_1^*\eta_1^*+2\text{i}\lambda_{1,I}\eta_1)-2\epsilon\lambda_{1,R}\lambda_1\eta_1^{*2},\\
&p_2=4\omega_2(\lambda_{1}\eta_{1}+\lambda_{1}^*\eta_{1}^*)-\epsilon(\lambda_1\eta_1^*+\lambda_1^*\eta_1)|\eta_1|^2,~~p_3=4\lambda_{1,R}^2|2\omega_2-\epsilon\eta_1^2|^2-32\epsilon\omega_2|\lambda_1|^2\eta_{1,I}^2,\\
&H_1=\lambda_{1,I}\text{Im}[\lambda_1\eta_1\text{e}^{2\text{i}(\chi_{1,I}-\xi_{1,I})}]+|\lambda_1|^2\eta_{1,R},H_2=|\lambda_1|^2(2\omega_2-\epsilon|\eta_1|^2)+\omega_1\lambda_{1,R}^2\text{e}^{-2(\chi_{1,R}-\xi_{1,R})}.
\end{aligned}\end{equation}
For the focusing gSS equation with $a=-1,b=\frac{1}{2}$, if set $\lambda_1=1+\frac{\text{i}}{2}$$(\lambda_{1,I}^2<3\lambda_{1,R}^2)$, this solution is bright-dark$(1,1)$ breather, which is the mixture of bright single-peak breather and dark single-peak breather(see Fig. 6$(a)$); if set $\lambda_1=\frac{1}{3}+\text{i}$$(\lambda_{1,I}^2>3\lambda_{1,R}^2)$, this solution is bright-dark$(1,2)$ breather, which is the mixture of one bright single-peak breather and dark double-peak breather(see Fig. 6$(b)$).

For the defocusing gSS equation with $a=\frac{1}{3},b=-1$, set $\lambda_1=1+\frac{\text{i}}{10}$, this solution is located dark breather solution(see Fig. 7$(a)$); set $\lambda_1=1+\frac{\text{i}}{2}$, this solution is bright-dark$(1,1)$ breather(see Fig. 7$(b)$); set $\lambda_1=1+2\text{i}$, this solution is bright-dark$(1,2)$ breather(see Fig. 7$(c)$). We can see that if fix the value of $\lambda_{1,R}$, the shape of the soliton from dark breather becomes into bright-dark breather with the increase of $\lambda_{1,I}$. The bright-dark breather is $(1,1)$ or $(1,2)$ depend on $3\lambda_{1,R}^2>\lambda_{1,I}^2$ or $3\lambda_{1,R}^2<\lambda_{1,I}^2$.

In particular, if $\lambda_1=\text{i}\beta_1$, when $2\epsilon\omega_2+\beta_1^2<0$, i.e. $\tau_j$ is a pure imaginary number, the solution (34) reduces into
\begin{eqnarray*}
u^{(1)}=1+\frac{4\epsilon\beta_1(2\omega_1\beta_1\text{cosh}(\chi_1)^2+\omega_1\tau_1\text{sinh}(2\chi_1)-2\omega_2\tau_1\text{cosh}(\chi_1)\text{e}^{-\xi_1})}
{\omega_1(2\omega_2+\epsilon\eta_1\kappa_1+(2\omega_2-\epsilon\eta_1^2)\text{cosh}(2\chi_1))+2\epsilon\tau_1(\omega_1\beta_1\text{e}^{-2\chi_1}+\omega_2\tau_1\text{e}^{-2\xi_1})}.
\end{eqnarray*}
This solution for the focusing gSS equation is periodic-like solution(see Fig. 8). This type of solution to SS equation has appeared in [5].

If $d_{11}=1,d_{12}=1,d_{13}=0$, we obtain the solution to the focusing gSS equation
\begin{equation}\label{solution3-11}\begin{aligned}
&u^{(1)}=1+\frac{16\lambda_{1,I}(Re[\Delta_1]+2|\lambda_1|^2Re[H_3\text{cosh}(\chi_1^*)](2(a+b)|\text{cosh}(\chi_1)|^2-|H_3|^2))}
{4|\lambda_1|^2(2(a+b)|\text{cosh}(\chi_1^*)|^2-|H_3|^2)^2+(\lambda_1-\lambda_1^*)^2\big|2(a+b)\text{cosh}(\chi_1^*)^2-H_3^2\big|^2},\\
&\Delta_1=\lambda_1(\lambda_1-\lambda_1^*)H_{1}\text{cosh}(\chi_1)(2(a+b)\text{cosh}(\chi_1)^2-H_1^{*2}),~~H_3=\tau_1\text{sinh}(\chi_1)-\text{i}\lambda_1\text{cosh}(\chi_1).
\end{aligned}\end{equation}
It is clear that $u^{(1)}$ given by (35) is a real solution.
Set $\lambda_1=1+\frac{\text{i}}{2}$, $|u^{(1)}|$$(\lambda_{1,I}^2<3\lambda_{1,R}^2)$ is a bright-bright breather, which is the mixture of two bright single-peak breather waves(see Fig. 9$(a)(c)$); Setting $\lambda_1=\frac{1}{3}+\text{i}$$(\lambda_{1,I}^2>3\lambda_{1,R}^2)$, $|u^{(1)}|$ is a bright-bright breather, which is the mixture of two bright breather(see Fig. 9$(b)(d)$).

If $d_{11}=1,d_{12}=1,d_{13}=1$, the solution of the focusing gSS equation can be written as
\begin{equation}\label{solution3-3}\begin{aligned}
&u^{(1)}=1+\frac{8\lambda_{1,I}((2\text{cosh}(\chi_1)+\text{e}^{\xi_1})\Delta_4+(2\text{cosh}(\chi_1^*)-\text{e}^{\xi_1^*})\Delta_4^*)}
{4|\lambda_1|^2(H_5-2|H_3|^2)^2-4\lambda_{1,I}^2\big|H_4-2H_3^{*2}\big|^2},\\
&\Delta_4=\lambda_1(\lambda_1-\lambda_1^*)H_3(H_4^*-2H_3^{*2})+2\lambda_1^*(H_5-2|H_3|^2),\\
&H_4=4(a+b)\text{cosh}(\chi_1)^2-(a-b)\text{e}^{2\xi_1},H_5=4(a+b)|\text{cosh}(\chi_1)|^2+(a-b)\text{e}^{2\xi_{1,R}}.
\end{aligned}\end{equation}
Setting $\lambda_1=\frac{1}{3}+\text{i}$, this solution is resonant $2$-breather solution(see Fig. 10$(a)$).
In particular, if $\lambda_1=\text{i}\beta_1$, the solution (36) reduces into
\begin{equation}\label{solution3-4}
u^{(1)}=-1+\frac{2\tau_1^2}{2(a+b)-\beta_1^2\text{cosh}(2\chi_1)-\beta_1\tau_1\text{sinh}(2\chi_1)},
\end{equation}
which is the exact solution for the mKdV equation. When $\beta_1^2+2(a+b)<0$, i.e. $\chi_1$ is imaginary, the solution (37) is periodic in space and time(see Fig. 10$(b)(c)$). Spatial period of this solution is $\frac{\pi}{2|a+b-\beta_1^2|\sqrt{-\beta_1^2-2(a+b)}}$, and time period of the solution is $\frac{\pi}{\sqrt{-\beta_1^2-2(a+b)}}$. When $\beta_1^2+2(a+b)>0$, i.e. $\chi_1$ is real, the solution (37) is hump-soliton(see Fig. 10$(d)$), the peak value of $|u^{(1)}|$ is $\big|\frac{2\tau_1^2}{2(a+b)-\beta_1\sqrt{|2(a+b)|}}-1\big|$ located in the line
\begin{eqnarray*}
x=-4(a+b-\beta_1^2)t+\frac{1}{4\sqrt{\beta_1^2+2(a+b)}}\ln\big|\frac{\beta_1-\sqrt{\beta_1^2+2(a+b)}}{\beta_1+\sqrt{\beta_1^2+2(a+b)}}\big|.
\end{eqnarray*}

\subsection{ $2$-soliton and breather solutions }

By using two-fold DT, we obtain interaction solution to gSS equation of soliton, breather, resonant interaction and periodic solution from nonzero seed solution.
For the focusing gSS equation, we derive $2$-breather solutions, which describe collision of bright-dark, bright-bright breather and periodic wave; For the defocusing gSS equation, we obtain $2$-breather solutions, displaying collision of two dark breather solitons. In order to facilitate the calculation, here we set $d_{11}=d_{21}=1$.

For focusing gSS equation, suppose $\lambda_{1,R}\lambda_{1,I}\lambda_{2,R}\lambda_{2,I}\neq0$, e.g. $\lambda_1=1+\frac{\text{i}}{2}$, $\lambda_2=\frac{1}{2}+\text{i}$, and set $d_{12}=d_{23}=0$, $d_{13}=d_{22}=1$, $2$-breather solution is obtained, which displays that a bright-dark$(1,1)$ breather $u_{16}^{-}$ and a bright-bright$(1,2)$ breather $u_{26}^{-}$ become into bright-dark$(1,1)$ breather $u_{16}^{+}$ and bright-dark$(1,2)$ breather $u_{26}^{+}$(see Fig. 11$(a)$).
The asymptotic behavior of the focusing Sasa-Satsuma equation is
\begin{eqnarray*}
u^{(2)}\longrightarrow \begin{cases}
u_{16}^{-}+u_{26}^{-},\quad t\rightarrow-\infty,\\
u_{16}^{+}+u_{26}^{+},\quad t\rightarrow+\infty;
\end{cases}
\end{eqnarray*}
where $u_{16}^{\pm},u_{26}^{\pm}$ are too complicated, and we omit them.

When $d_{12}=d_{22}=0$, $d_{13}=d_{23}=0$, the $2$-breather solution shows the interaction of two bright-dark breathers(see Fig. 11$(b)$), and its asymptotic behavior can be also obtained by the asymptotic analysis. When $d_{12}=0$, $d_{13}=d_{22}=d_{23}=1$, the $2$-breather solution describes the interaction of bright-dark breather and resonant$(2,1)$-breather(see Fig. 11$(c)$).

Let us present some new solutions to Sasa-Satsuma equation(3). Set $\lambda_1=1-\frac{\text{i}}{2},\lambda_2=\frac{1}{2}-\frac{\text{i}}{2}$, $d_{13}=d_{23}=1$ and $d_{12}=d_{22}=0$. The $2$-breather solution displays the collision of two dark breathers for defocusing Sasa-Satsuma equation(see Fig. 11$(d)$).

For focusing SS equation, we set $\lambda_1=1+\frac{\text{i}}{2}$ and $\lambda_2=\text{i}\beta_2$ ($\beta_2^2+2(a+b)<0$). When $d_{12}=d_{22}=0$ and $d_{13}=d_{23}=1$, this solution is the mixture of bright-dark breather and periodic-like wave(see Fig. 12$(a)(b)$); when $d_{13}=d_{22}=0$ and $d_{12}=d_{23}=1$, this solution is the mixture of bright-bright breather and periodic-like wave(see Fig. 12$(c)(d)$). When $d_{12}=0$, and $d_{13}=d_{22}=d_{23}=1$, if $\beta_2^2+2(a+b)<0$, this solution shows the interaction of a bright-dark$(1,1)$ breather with periodic wave(see Fig. 13$(a)$); if $\beta_2^2+2(a+b)>0$, this solution describes the collision of a bright-dark$(1,1)$ breather with hump soliton(see Fig. 13$(b)$). If set $d_{12}=0$, $d_{13}=d_{22}=d_{23}=1$, and $\beta_2^2+2(a+b)<0$, this solution displays the interaction of a bright-bright$(1,1)$ breather with periodic wave; if $\beta_2^2+2(a+b)>0$, this solution presents the collision of a bright-dark$(1,1)$ breather with hump soliton. When $\lambda_1=\text{i},\lambda_2=\frac{\text{i}}{2}$, $d_{12}=d_{13}=d_{22}=d_{23}=1$, this solution is bright-bright breather-periodic solution(see Fig. 13$(c)$). When $\lambda_1=\frac{9\text{i}}{5},\lambda_2=\frac{8\text{i}}{5}$, $d_{12}=d_{13}=d_{22}=d_{23}=1$, this solution depicts the interaction of hump soliton and W-hump soliton(see Fig. 13$(d)$).


\section{Conservation laws}

The existence of infinite number conservation laws is an important embodiment of the integrability of the equation. In this section, we discuss infinitely number conservation laws for the gSS equation by using the Lax representation. Suppose $\Psi=(\psi_1,\psi_2,\psi_3)^T$ is an eigenfunction of the eigenvalue problem(6). Let
\begin{eqnarray*}
\Phi_1=\frac{\psi_1(x,t,\lambda)}{\psi_3(x,t,\lambda)},~~~\Phi_2=\frac{\psi_2(x,t,\lambda)}{\psi_3(x,t,\lambda)}.
\end{eqnarray*}
Then the eigenvalue problem(6) can be written as the Reccati equation
\begin{equation}\label{RE}\begin{aligned}
&\Phi_{1,x}=u+2\text{i}\lambda\Phi_1-\epsilon(au^*+bu)\Phi_1^2-(au+bu^*)\Phi_1\Phi_2,\\ \nonumber
&\Phi_{2,x}=\epsilon u^*+2\text{i}\lambda\Phi_2-\epsilon(au^*+bu)\Phi_1\Phi_2-(au+bu^*)\Phi_2^2.
\end{aligned}\end{equation}
Expanding $\Phi_1$ and $\Phi_2$ as
\begin{eqnarray*}
\Phi_{1}=\sum_{n=1}^{\infty}C_n^{(1)}\lambda^{-n},~~~\Phi_{2}=\sum_{n=1}^{\infty}C_n^{(2)}\lambda^{-n}.
\end{eqnarray*}
and substituting $\Phi_1$ and $\Phi_2$ into the Reccati equation (38), we obtain
\begin{equation}\label{CC}\begin{aligned}
&C_1^{(1)}=\frac{\text{i}}{2}u,~~~C_2^{(1)}=\frac{1}{4}u_x,~~~C_3^{(1)}=-\frac{\text{i}}{8}u_{xx}+\frac{\text{i}}{4}\epsilon a|u|^2u+\frac{\text{i}}{8}\epsilon bu(u^2+u^{*2}),\\
&C_1^{(2)}=\frac{\text{i}}{2}\epsilon u^*,~~~C_2^{(2)}=\frac{1}{4}\epsilon u^*_x,~~~C_3^{(2)}=-\frac{\text{i}}{8}\epsilon u^*_{xx}+\frac{\text{i}}{4} a|u|^2u^*+\frac{\text{i}}{8} bu^*(u^2+u^{*2}),
\end{aligned}\end{equation}
and the recursion relation
\begin{eqnarray*}
&&C_{k+1}^{(1)}=-\frac{\text{i}}{2}\left(C_{k,x}^{(1)}+\sum_{j=1}^{k-1}(\epsilon(au^*+bu)C_j^{(1)}C_{k-j}^{(1)}+(au+bu^*)C_j^{(1)}C_{k-j}^{(2)})\right),\\ \\
&&C_{k+1}^{(2)}=-\frac{\text{i}}{2}\left(C_{k,x}^{(2)}+\sum_{j=1}^{k-1}(\epsilon(au^*+bu)C_j^{(2)}C_{k-j}^{(1)}+(au+bu^*)C_j^{(2)}C_{k-j}^{(2)})\right),k=3,4,\cdots. \end{eqnarray*}
According to the compatibility condition $(\ln\psi_3)_{xt}=(\ln\psi_3)_{tx}$, we have $\frac{\partial}{\partial t}\mathcal{P}=\frac{\partial}{\partial x}\mathcal{J}$, where
\begin{eqnarray*}
&&\mathcal{P}=\epsilon(au^*+bu)\Phi_1+(au+bu^*)\Phi_2,\\
&&\mathcal{J}=-2\text{i}\epsilon\lambda(2a|u|^2+bu^2+bu^{*2})+\epsilon\Phi_1(4\lambda^2(au^*+bu)+2\text{i}\lambda(au_x^*+bu_x)+4\epsilon a^2|u|^2u^*\\
&&~+2\epsilon b^2(u^3+|u|^2u^*)+2ab\epsilon(3|u|^2u+u^{*3})-au^*_{xx}-bu_{xx})+\Phi_2(4\lambda^2(au+bu^*)\\
&&~+2\text{i}\lambda(au_x+bu_x^*)+4\epsilon a^2|u|^2u+2\epsilon b^2(u^{*3}+|u|^2u)+2ab\epsilon(3|u|^2u^*+u^{3})-au_{xx}-bu^*_{xx}).
\end{eqnarray*}
Then expanding $\mathcal{P}$ and $\mathcal{J}$ in the form
\begin{eqnarray*}
\mathcal{P}=\sum_{k=1}^{\infty}\mathcal{P}_k\lambda^{-n},~~~\mathcal{J}=\sum_{k=1}^{\infty}\mathcal{J}_k\lambda^{-n},
\end{eqnarray*}
yields infinite number of conservation laws $\frac{\partial}{\partial t}\mathcal{P}_{k}=\frac{\partial}{\partial x}\mathcal{J}_{k}$, where
\begin{eqnarray*}
&&\mathcal{P}_1=\frac{\text{i}}{2}\epsilon(2a|u|^2+bu^2+bu^{*2}),\\
&&\mathcal{P}_2=\frac{1}{8}\epsilon(2a|u|^2+bu^2+bu^{*2})_x,\\
&&\mathcal{P}_3=-\frac{\text{i}}{8}\epsilon\left(a(u_{xx}u^*+uu^*_{xx})+b(uu_{xx}+u^*_{xx})-\epsilon(2a|u|^2+bu^2+bu^{*2})^2\right),\\
&&\mathcal{P}_k=\epsilon C_k^{(1)}(au^*+bu)+C_k^{(2)}(au+bu^*),k=4,5,\cdots,
\end{eqnarray*}
and
\begin{eqnarray*}
&&\mathcal{J}_1=\frac{\text{i}}{2}\epsilon(-2a(u_{xx}u^*-u_xu_x^*+uu^*_{xx})+b(u_x^2-2uu_{xx}+u^{*2}_x-2u^*u^*_{xx})\\
&&~~~+12\epsilon a^2|u|^4+3\epsilon b^2(u^2+u^{*2})^2+12\epsilon ab|u|^2(u^2+u^{*2})),\\
&&\mathcal{J}_2=\frac{1}{4}\epsilon(-a(u_{xxx}u^*+uu^*_{xxx})-b(uu_{xxx}+u^*u^*_{xxx})+12\epsilon a^2|u|^2(uu^*_x+u^*u_x)\\
&&~~~+6\epsilon b^2(u^2+u^{*2})(uu_x+u^*u^*_x)+6\epsilon ab(3|u|^2(uu_x+u^*u^*_x)+u^{*3}u_x+u^3u^*_x)),\\
&&\mathcal{J}_j=C_j^{(1)}(4a^2|u|^2u^*+2b^2u(u^2+u^{*2})+2abu^*(3u^2+u^{*2})-\epsilon au^*_{xx}-b u_{xx})\\
&&~~~+\epsilon C_j^{(2)}(4a^2|u|^2u+2b^2u^*(u^2+u^{*2})+2abu(3u^{*2}+u^{2})-\epsilon au_{xx}-b u^*_{xx})\\
&&~~~+2\text{i}\epsilon C_{j+1}^{(1)}(au^*_x+bu_x)+2\text{i}\epsilon C_{j+1}^{(2)}(au_x+bu^*_x)\\
&&~~~+4\epsilon C_{j+2}^{(1)}(au^*+bu)+4\epsilon C_{j+2}^{(2)}(au+bu^*),j=3,4,\cdots.
\end{eqnarray*}

It is easy to find that
\begin{equation*}
(\text{ln}\psi_3)_x=-\text{i}\lambda+\epsilon(au^*+bu)\Phi_1+(au+bu^*)\Phi_2,
\end{equation*}
then $\epsilon(au^*+bu)C_k^{(1)}+(au+bu^*)C_k^{(2)}$ would be the density of the conservation law, and we obtain an infinite number of conserved quantities
\begin{eqnarray*}
I_k=\int_{-\infty}^{\infty}(-1)^{k+1}(2\text{i})^k\epsilon(\epsilon(au^*+bu)C_k^{(1)}+(au+bu^*)C_k^{(2)})dx,~~k=1,2,\ldots.
\end{eqnarray*}
Substituting (38) into above equation, we can derive the conserved quantities, where the first four conserved quantities are
\begin{eqnarray*}
&&I_1=\int_{-\infty}^{\infty}-(2a|u|^2+b(u^2+u^{*2}))dx,~~~I_2=0,\\
&&I_3=\int_{-\infty}^{\infty}(-a(uu^*_{xx}+u^*u_{xx})-b(uu_{xx}+u^*u^*_{xx})+2a|u|^2+b(u^2+u^{*2})^2)dx,\\
&&I_4=\int_{-\infty}^{\infty}(a(uu^*_{xxx}+u^*u_{xxx})+b(uu_{xxx}+u^*u^*_{xxx}))dx.
\end{eqnarray*}

\section{Conclusion}
In this paper, we have constructed $N$-fold DT of the gSS equation. We have seen that the construction of DT for gSS equation is difficult. By using the DT, various of soliton solutions for the focusing and defocusing gSS equation with zero and nonzero seed solution have been derived, including hump-soliton solution, breather-type solution, resonant $2$-breather solution, periodic solution. Furthermore, dynamics properties and asymptotic behavior of these solutions have been analyzed. Compare with soliton solutions discussed in[16], soliton solutions derived by the zero seed solution are agree with Eq.(3.23) in [16], while soliton solution with non-zero seed of the gSS equation was not discussed in [16]. Compared with the research results of the Sasa-Satsuma equation in the literatures, we found several novel soliton solutions, including breather-like, resonant $2$-breather solution, and the interaction solution of bright-bright breather and other type solitons. By solving the related Riccati equation, we have derived the infinite number conservation laws and conserved quantities for the gSS equation.

\section*{Acknowledgements}
The work of ZNZ is supported by National Natural Science Foundation of China under Grant No.12071286, and by the Ministry of Economy and Competitiveness of Spain under contract PID2020-115273GB-I00 (AEI/FEDER,EU).


\begin{thebibliography}{60}
\bibitem{Sasa(1991)}
Sasa N, Satsuma J. 1991 New-type of soliton solutions for a higher-order nonlinear Schr\"{o}dinger equation. J. Phys. Soc. Jpn. 60, 409--417.(doi:org/10.1143/JPSJ.60.409)
\bibitem{Mihalache(1993)}
Mihalache D, Torner L, Moldoveanu F, Panoiu N-C, Truta N. 1993 Inverse-scattering approach to femtosecond solitons in monomode optical fibers. Phys. Rev. E. 48, 4699--4709.(doi:10.1103/PhysRevE.48.4699)

\bibitem{Mihalache(1993a)}
Mihalache D, Torner L, Moldoveanu F, Panoiu N-C, Truta N. 1993 Soliton solutions for a perturbed nonlinear Schr\"{o}dinger equation. J. Phys. A: Math. Gen. 26, L757-L765.(doi:10.1088/0305-4470/26/17/001)

\bibitem{Xu(2014)}
Xu T, Wang D-H, Li M, Liang H. 2014 Soliton and breather solutions of the Sasa-Satsuma equation via the Darboux transformation. Phys. Scr. 89, 075207.(doi:10.1088/0031-8949/89/7/075207)

\bibitem{Nimmo(2015)}
Nimmo J-J-C, Yilmaz H. 2015 Binary Darboux transformation for the Sasa-Satsuma equation. J. Phys. A: Math. Theor. 48, 425202.(doi:10.1088/1751-8113/48/42/425202)

\bibitem{Xu(2015)}
Xu T, Li M, Li L. 2015 Anti-dark and Mexican-hat solitons in the Sasa-Satsuma equation on the continuous wave background. EPL. 109, 30006.(doi:10.1209/0295-5075/109/30006)
\bibitem{Gilson(2003)}
Gilson C, Hietarinta J, Nimmo J, Ohta Y. 2002 Sasa-Satsuma higher-order nonlinear Schr\"{o}dinger equation and its bilinearization and multisoliton solutions. Phys. Rev. E. 68 016614.(doi:10.1103/PhysRevE.68.016614)
\bibitem{Bandelow(2012)}
Bandelow U, Akhmediev N. 2012 Sasa-Satsuma equation: Soliton on a background and its limiting cases. Phys. Rev. E. 86, 026606.(doi:10.1103/PhysRevE.86.026606)
\bibitem{Zhao(2014)}
Zhao L-C, Li S-C, Ling L-M. 2014 Rational W-shaped solitons on a continuous-wave background in the Sasa-Satsuma equation. Phys. Rev. E. 89, 023210.(doi:10.1103/PhysRevE.89.023210)
\bibitem{Ohta(2010)}
Ohta Y. 2010 Dark soliton solution of Sasa-Satsuma equation. AIP Conference Proceedings, 1212, 114--121.(doi:org/10.1063/1.3367022)
\bibitem{Ghosh(1999)}
Ghosh S, Kundu A, Nandy S. 1999 Soliton solutions, Liouville integrability and gauge equivalence of Sasa Satsuma equation. J. Math. Phys. 40, 1993--2000.(doi:org/10.1063/1.532845)
\bibitem{Xu(2013)}
Xu J, Fan E-G. 2013 The unified transform method for the Sasa-Satsuma equation on the half-line. Proc. R. Soc. A. 469, 20130068.(doi:org/10.1098/rspa.2013.0068)
\bibitem{Xu(2018)}
Xu J, Zhu Q-Z, Fan E-G. 2018 The initial-boundary value problem for the Sasa-Satsuma equation on a finite interval via the Fokas method. J. Math. Phys. 59, 073508.(doi:org/10.1063/1.5047140)
\bibitem{LiuH(2018)}
Liu H, Geng X-G, Xue B. 2018 The Deift-Zhou steepest descent method to long-time asymptotics for the Sasa-Satsuma equation. J. Differ. Equ. 265, 5984--6008.(doi:org/10.1016/j.jde.2018.07.026)
\bibitem{LiuN(2019)}
Liu N, Guo B-L. 2019 Long-time asymptotics for the Sasa-Satsuma equation via nonlinear steepest descent method. J. Math. Phys. 60, 011504.(doi:org/10.1063/1.5061793)
\bibitem{Geng(2016)}
Geng X-G, Wu J-P. 2016 Riemann-Hilbert approach and N-soliton solutions for a generalized Sasa-Satsuma equation. Wave Motion, 60, 62--72.(doi:org/10.1016/j.wavemoti.2015.09.003)
\bibitem{Geng(2020)}
Wang K-D, Geng X-G, Chen M-M, Li R-M. 2020 Long-time asymptotics for the generalized Sasa-Satsuma equation. AIMS Mathematics. 5, 7413--7437.(doi: 10.3934/math.2020475)
\end{thebibliography}
\end{document}